\documentclass[%
 aip,
 amsmath,amssymb,
reprint,%
]{revtex4-1}
\usepackage{scrextend}
\usepackage{lineno}

\usepackage{graphicx}
\usepackage{dcolumn}
\usepackage{bm}
\usepackage{amsmath}
\usepackage{multirow}

\newcommand{\mean}[1]{\mbox{$\langle{#1}\rangle$}}

\usepackage[colorlinks=true]{hyperref}
\usepackage{amsmath}
\usepackage{amssymb}
\usepackage{array,booktabs}
\usepackage{graphicx}
\usepackage{siunitx}
\usepackage{amsbsy}

\begin{document}
\title{A Field-Enhanced Conduction-Cooled Superconducting Cavity for High-Repetition-Rate Ultrafast Electron Bunch Generation}

\author{O. Mohsen} 
\affiliation{Northern Illinois Center for Accelerator \& Detector Development and Department of Physics, Northern Illinois University, DeKalb IL 60115, USA}
\author{D. Mihalcea}
\affiliation{Northern Illinois Center for Accelerator \& Detector Development and Department of Physics, Northern Illinois University, DeKalb IL 60115, USA}
\author{N. Tom}
\affiliation{Northern Illinois Center for Accelerator \& Detector Development and Department of Mechanical Engineering, Northern Illinois University, DeKalb IL 60115, USA}
\author{N. Adams}
\affiliation{Northern Illinois Center for Accelerator \& Detector Development and Department of Mechanical Engineering, Northern Illinois University, DeKalb IL 60115, USA}
\author{R. C. Dhuley}
\affiliation{Fermi National Accelerator Laboratory, Batavia, IL 60510, USA}
\author{M. I. Geelhoed}
\affiliation{Fermi National Accelerator Laboratory, Batavia, IL 60510, USA}
\author{A. McKeown} 
\affiliation{Northern Illinois Center for Accelerator \& Detector Development and Department of Mechanical Engineering, Northern Illinois University, DeKalb IL 60115, USA}
\author{V. Korampally}
\affiliation{Northern Illinois Center for Accelerator \& Detector Development and Department of Electrical Engineering, Northern Illinois University, DeKalb IL 60115,~USA}
\author{P.  Piot}
\affiliation{Northern Illinois Center for Accelerator \& Detector Development and Department of Physics, Northern Illinois University, DeKalb IL 60115, USA}
\affiliation{Fermi National Accelerator Laboratory, Batavia, IL 60510, USA}
\author{I. Salehinia}
\affiliation{Northern Illinois Center for Accelerator \& Detector Development and Department of Mechanical Engineering, Northern Illinois University, DeKalb IL 60115, USA}
\author{J. C. T. Thangaraj}
\affiliation{Fermi National Accelerator Laboratory, Batavia, IL 60510, USA}
\author{T. Xu}
\affiliation{Northern Illinois Center for Accelerator \& Detector Development and Department of Physics, Northern Illinois University, DeKalb IL 60115, USA}
%
\begin{abstract}
High-repetition-rate sources of bright electron bunches have a wide range of applications. They can directly be employed as probes in electron-scattering setups, or serve as a backbone for the generation of radiation over a broad range of the electromagnetic spectrum. This paper describes the development of a compact sub-Mega-electronvolt (sub-MeV) electron-source setup capable of operating at MHz repetition rates and forming sub-picosecond electron bunches with transverse emittance below 20~nm. The setup relies on a conduction-cooled superconducting single-cell resonator with its geometry altered to enhance the field at the surface of the emitter. The system is designed to accommodate cooling using a model a $2$~W at 4.2 K pulse tube cryogen-free cryocooler. Although we focus on the case of a photoemitted electron bunch, the scheme could be adapted to other emission mechanisms. 
\end{abstract}

\date{\today}

\maketitle

\section{introduction}
Since its first application to particle accelerator decades ago, the superconducting radio-frequency (SRF) technology has steadily improved and enabled a new class of efficient accelerators with an increasing number of applications in fundamental science at national laboratories, e.g. in nuclear physics~\cite{cebaf}, and free-electron lasers~\cite{irdemo,ttf1}. One of the main challenges limiting the dissemination of SRF-based accelerators is the complex infrastructure associated with the required cryogenic system (e.g. liquefaction and storing of Helium) necessary to maintain the SRF resonators below the critical temperature~\cite{Kephart}. Most recently, cryogen-free cooling techniques applied to SRF techniques have demonstrated the operation of superconducting resonators with high-quality factors $Q \sim {\cal O}(10^{9})$ and accelerating gradient $E_{acc} \sim 7 $~MV/m~\cite{Ram2020}. Cryogen-free systems generally employ compact closed-cycle cryocoolers contacted to the cavity via thermal-conduction links~\cite{RamLinks}. Such an achievement could eventually enable the dissemination of compact efficient SRF based accelerator with an array of societal applications. Likewise, a conduction-cooled SRF resonator coupled with an electron-emission scheme could facilitate the generation of bright ultrafast electron beams at high-repetition rates that could be disseminated as a research tool or for societal applications.

This paper discusses the design of a conduction-cooled SRF electron source capable of generating ultra-fast electron bunch with quality on par with state-of-the-art setups. The proposed source is based on a simple modification of a single-cell resonant cavity and can operate at MHz repetition rates. It is expected to have an array of applications in industry~\cite{jlabApplication,waterTreatment}, medicine~\cite{brainScan2013} and ultra-fast electron scattering~\cite{UED2015}. \\

\begin{figure}[t]
   \centering
   \includegraphics[width=0.995\columnwidth]{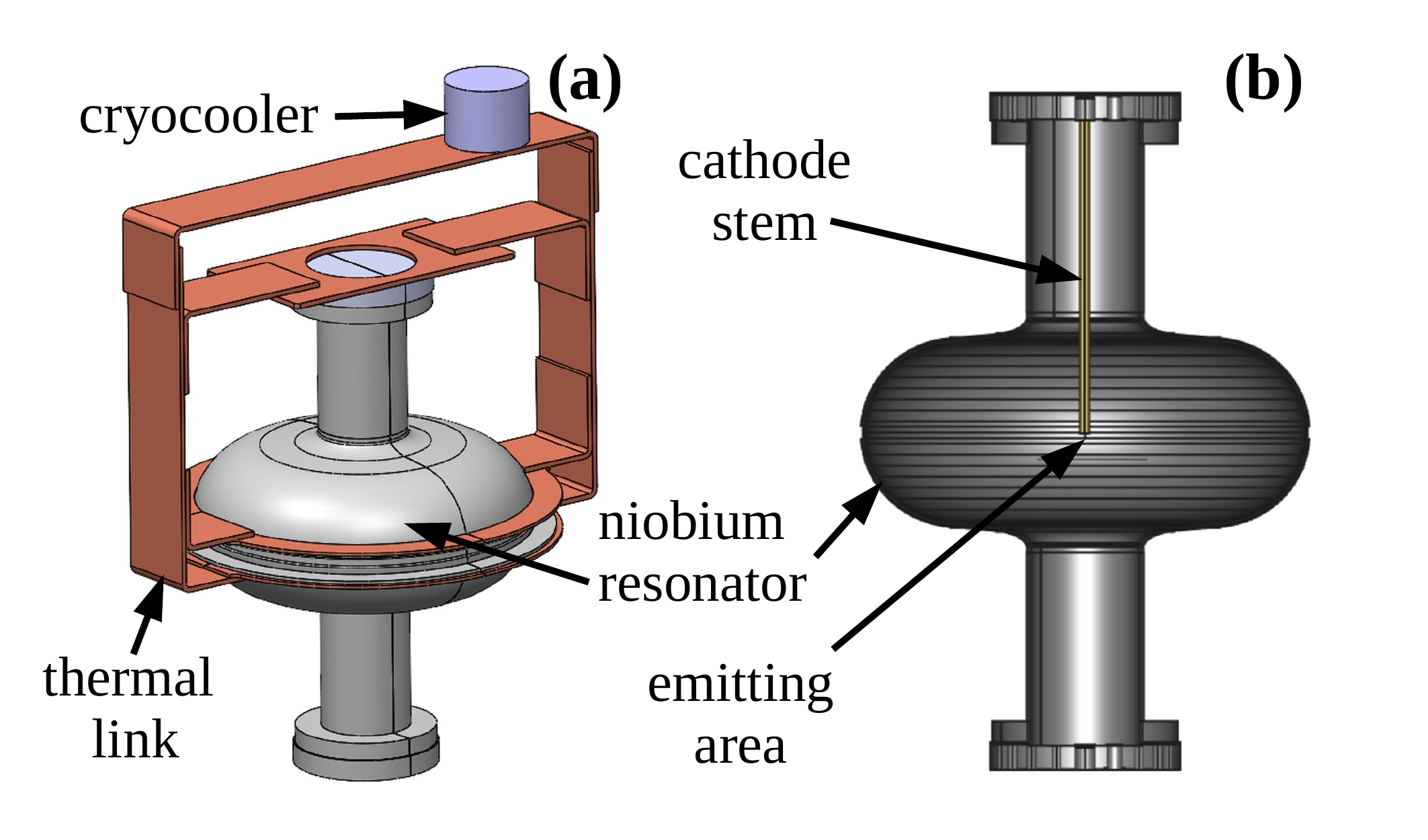}
   \caption{3D-rendition (a) and sectional view (b) of the modified single-cell 650-MHz superconducting resonator.}
   \label{fig:overview}
\end{figure}
The proposed electron-source concept appears in Fig.~\ref{fig:overview}: the system consists of a single-cell 650-MHz SRF cavity contacted to a cryocooler~\cite{4kcryo} via a network of thermal-conduction links made of high-purity aluminum \cite{RamLinks}. The setup is housed in a vacuum vessel that provides a thermally-insulating vacuum. The cavity, evacuated to ultra-high-vacuum pressure, is modified to incorporate a stem with a cathode (photoemission material) located at its extremity. The effect of the stem is twofold. First, it locally enhances the electric field at the emitting surface~\cite{Daoud} to support high-brightness electron emission given the favorable brightness scaling with the applied electric field at the cathode $E_c$ surface as ${\cal B} \propto E_c^{\alpha}$ where $\alpha\ge 1$ depends on the bunch initial aspect ratio~\cite{bazarov,daniele}. Second, the stem improves the transit time associated with the cavity as the shorter gap and higher electric field on the cathode surface significantly reduce the phase slippage between the emitted bunches and the oscillating electromagnetic fields.  

Given the change in the electromagnetic-field distribution introduced by the stem, the magnetic fields on its surface result in significant ohmic losses which limit the resonator maximum accelerating field. The compromises between performance and engineering design of the resonator are examined in this paper.
\section{expected radiofrequency properties}
\begin{figure}[h!]
   \centering
   \includegraphics[width=.925\columnwidth]{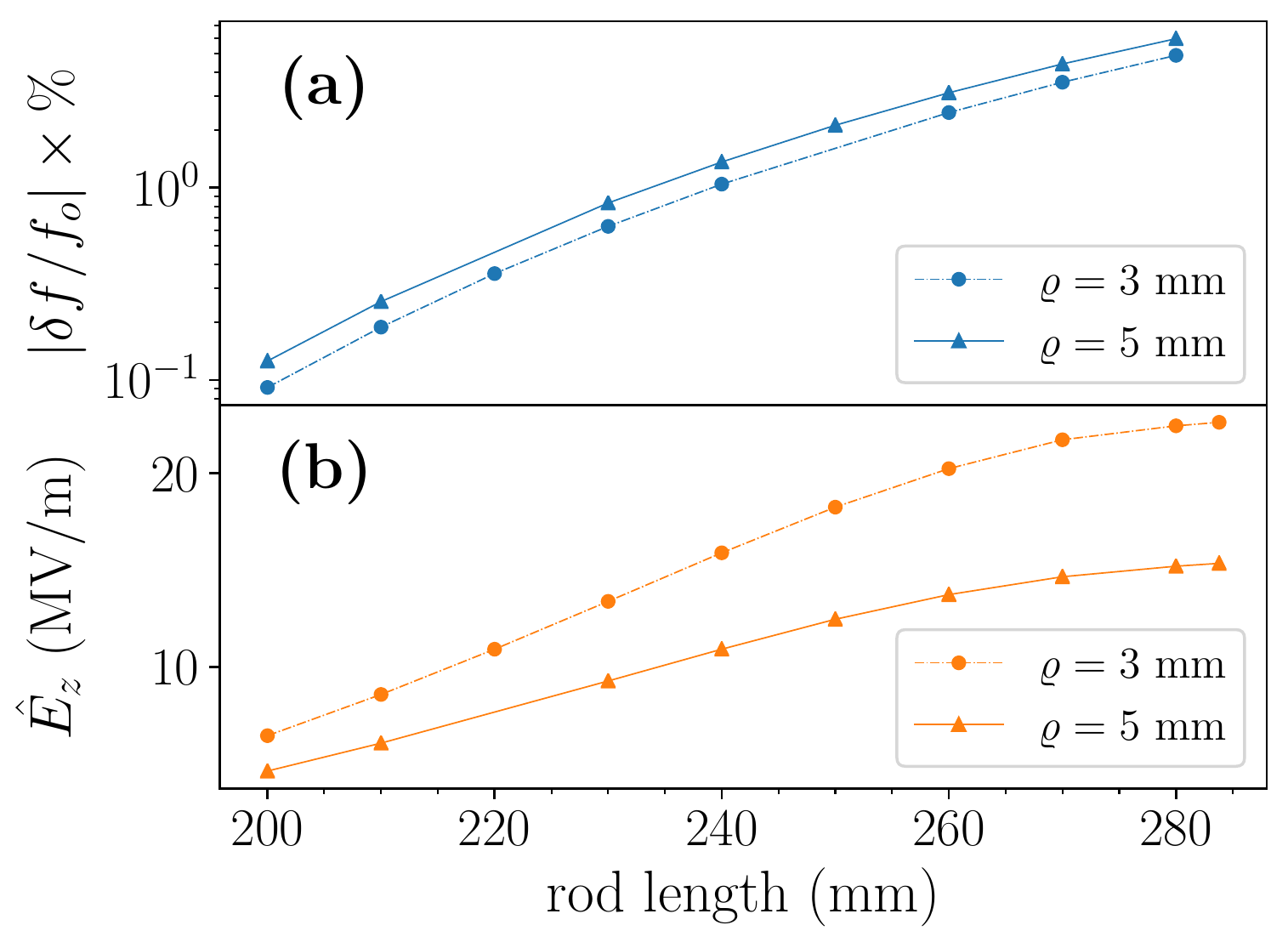}
   \caption{Relative shift in the resonant frequency (a) and maximum axial electric field $\hat{E_z}$ as a function of the stem length and two radii $\varrho$ (b).}
   \label{fig:rodimpact}
\end{figure}

In order to showcase the main advantages of the proposed concept we first investigate RF performances associated with the addition of a right-cylinder stem aligned along the cavity axis and with variable length. As expected, the stem impacts the resonant frequency of the cavity and leads to a significant enhancement of the electric-field amplitude at its extremity; see  Fig.~\ref{fig:rodimpact}. Furthermore, the field can be further enhanced by reducing the stem radius from 5 to 3-mm. The simulations indicate that despite the modest dissipated-power ($\sim 1$~W budget available), peak electric fields of $E_0\sim 20$~MV/m can be attained at the cathode surface. These field values are higher than state-of-the-art DC guns with similar complexity~\cite{doi:10.1063/1.4789395} and comparable to state-of-the-art CW RF gun~\cite{SANNIBALE201410}. The field-enhancement tapers off as the rod extremity reaches the cavity center (corresponding to a stem length $L=283$~mm) corresponding to the maximum attainable cathode-surface electric field. The corresponding relative frequency shift $\delta f/f$ remains below $\sim7$~\%. Such a change is irrelevant in practice as the resonator is powered by a CW solid-state amplifier (SSA) which can be specified to accommodate a given frequency. 

\begin{figure}[h!]
   \centering
   \includegraphics[width=1.0\columnwidth]{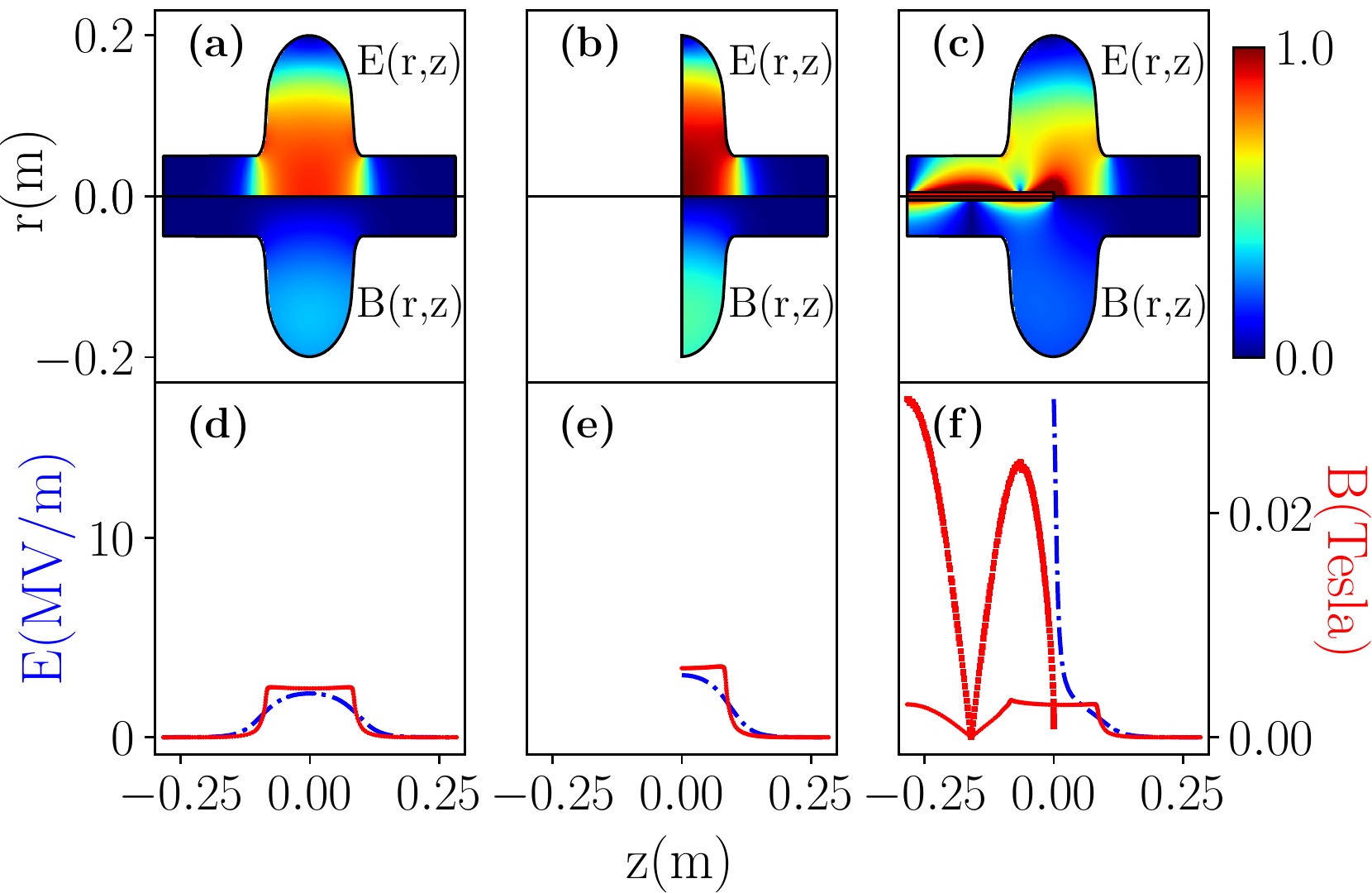}
   \caption{Geometry and associated peak-normalized electric ($r>0$) and magnetic ($r<0$) fields associated with the Full-cell (a), half-cell (b)  and modified field-enhanced resonator (c). The corresponding on-axis axial electric field $E_z(r=0,z)$ (blue) and surface magnetic field $B(\theta=0,z)$ (red) are respectively displayed in (d),(e) and (f). In (f) the thin (resp. thick) red trace corresponds to the surface magnetic field on the cavity-wall (resp. rod) surface. }
   \label{fig:fieldpattern}
\end{figure}

Consequently, to fully capitalize on the geometric enhancement, we focus on the case when the rod extremity is located at the center of the cavity. This configuration is generically denoted by ''FE". The resulting field distribution is compared with the two other configurations in Fig.~\ref{fig:fieldpattern}, full cavity with no stem ''FC" and half cell with no stem ''HC". The relevant RF parameters including the unloaded resonator quality factor $Q\equiv \omega \frac{U_{\Omega}}{P_{\Omega}}$ are directly evaluated with \textsc{omega3p} considering the properties of Nb at 5~K; see Tab.~\ref{table:compare}.
\begin{table}[h!]
\caption{RF parameters simulated for the three cavity configurations under consideration. }
\begin{ruledtabular}
\begin{tabular}{l  c  c c  c }
parameter & unit & FC & HC & FE \\
\hline
$f$   & MHz &  650 & 650 & 607 \\ 
$\hat{E}_z(r=0)$ & MV/m & 2.2 & 3.1 & 16.9 \\
${ U}$ & J& 0.123 & 0.123 & 0.080 \\ 
$Q$   & $-$ &3.12$\times 10^8$ & 3.12$\times 10^8$ &  1.8$\times 10^8$\\
\end{tabular}
\end{ruledtabular}
\label{table:compare}
\end{table}
Assuming the same dissipated power, the HC geometry gives a maximum field a factor $\sqrt{2}$ larger than the FC geometry.  For the adopted stem geometry, the FE model produces maximum electric field  $\sim 16$~MV/m. The associated magnetic-field pattern indicate substantially larger surface magnetic field of $\mbox{max}(B)\simeq 29.34$~mT; see  Fig.~\ref{fig:rodimpact}(f) which significantly increases the Ohmic losses. Finally, it should be pointed that for the simulated stored energy the CW power associated with the SSA is $P_f\simeq U\times f \sim 100$~W. For the photoemission process considered latter in this paper, the beam power is $P_b=q\times \nu \times {\cal E} \simeq 4 $~W$\ll P_f$ for a final beam energy ${\cal E}= 500$~keV, a single-bunch charge of $q=100$~fC with repetition frequency  $\nu= 80$~MHz. \\

\section{optimization of cathode stalk}

\begin{figure}[h!]
   \centering
   \includegraphics[width=.95\columnwidth]{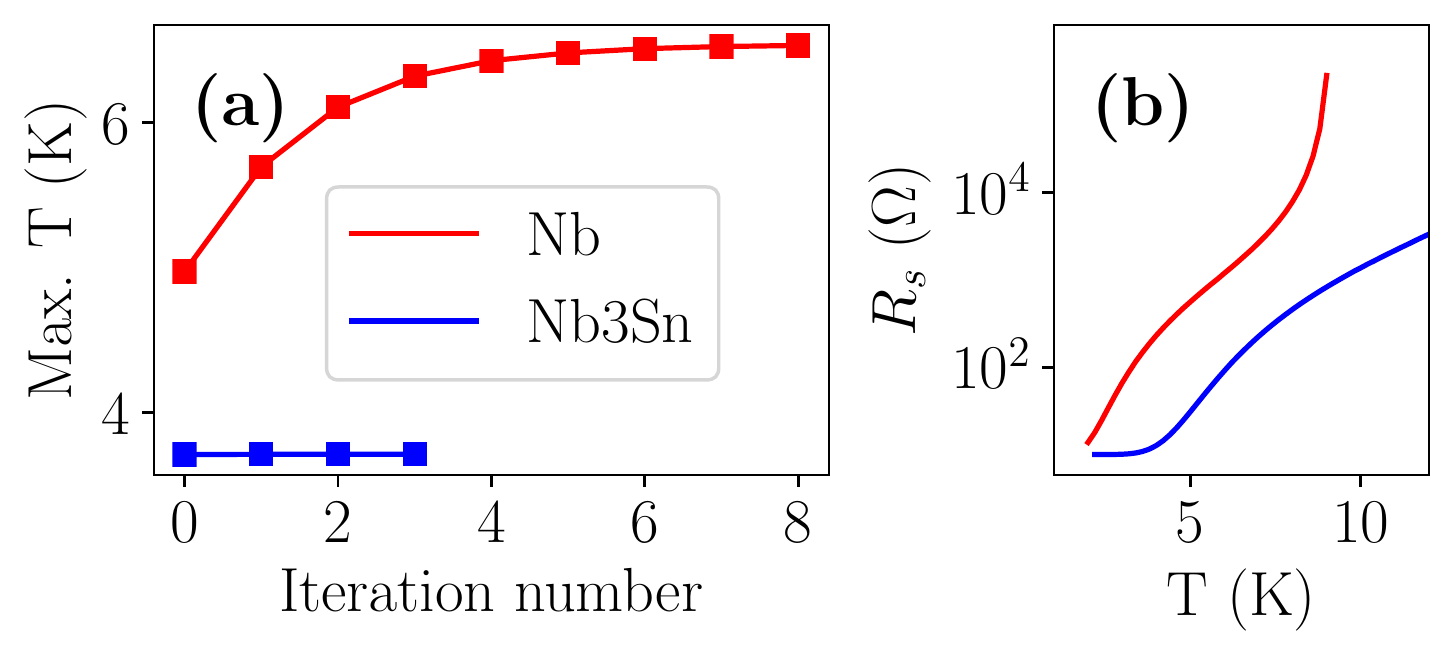}
   \caption{Evolution of the maximum temperature at the cathode-rod extremity during the iterative simulations (a) and temperature dependence of the surface resistance $R_s$ (b) used in the thermal simulation. The red and blue traces correspond respectively to the case of Nb (red) and Nb$_3$Sn. }
   \label{fig:temp_final}
\end{figure}

The surface magnetic field on the introduced cathode stem yields a 40\% reduction of the quality factor; see  Tab.~\ref{table:compare}. Most importantly, large surface magnetic fields can impact the superconducting state of the cavity. This is mainly when large magnetic fields result in localized heating of the rod surface that results in its temperature surpassing the critical temperature $T_c$. In the case of the cylindrical-rod geometry considered above, a thermal analysis using the field configuration display in Fig.~\ref{fig:fieldpattern}(c) indicates that the temperature can locally attain values $T>T_c=9.2$~K.  Consequently, we investigated possible approaches to reduce the equilibrium temperature to be below $T_c$ on the cavity wall and rod.  A thermal model coupled to the RF simulations was developed. The thermal simulation was implemented in the finite-element-analysis program {\sc ansys:workbench 19.2} and considers the system depicted in Fig.~\ref{fig:overview}(a) composed of the resonator, thermal links, and cryocooler. We assume the setup to be in thermal steady-state equilibrium where the dissipated power equates to the loading power of the cryocooler $P_k$. A constant-temperature ($T_k$) boundary condition was applied to the contact surface between the cryocooler and thermal link. 
Additionally, several discretized heat-flux boundary conditions were applied to the surfaces of the cavity and cathode mount to model the dissipated power. The dissipated power per unit area on the surfaces exposed to the electromagnetic field given by 
\begin{eqnarray}
\frac{dP}{dA}=\frac{1}{2}R_s |\pmb H|^2
\label{powerflux}
\end{eqnarray}
was numerically computed using the magnetic field distribution on the cavity internal surfaces simulated with  {\sc omega3p} and the surface resistance $R_S$ calculated with the algorithm \textsc{srimp}~\cite{SRIMP} which is based on quantum BCS theory of superconductors~\cite{BCScal}.
Given the dependence of the surface resistance on the temperature an iterative method is used. First, the dissipated power $\frac{dP^{(0)}}{dA}$ is evaluated from the electromagnetic simulations with {\sc omega3p} and used to compute the thermal heat flux in the thermal model for all the surfaces exposed to the electromagnetic field. The resulting temperature $T^{(0)}$ from the thermal simulation is then used to compute the dissipated power, taking into account the dependence $R_s(T)$, following the recursive relation (for $n=0$) $\frac{dP^{(n+1)}}{dA}=\frac{1}{2}R_s(T^{(n)}) |\pmb H_{\parallel}|^2$ 
The above process is repeated (for $n>0$) until convergence is attained. In addition to converging, it is also imperative that the maximum temperature $T_{max}$ (which is always at the rod tip) remains below the critical temperature $T_{c}$. Therefore, it is convenient to define the tolerance parameter in terms of $T_{max}$
\begin{eqnarray}
e_n \equiv \big| T_{max}^{(n)} -T_{max}^{(n-1)}\big|, 
\end{eqnarray}
with the condition $T_{max}< T_c$. It should be noted that since the heat flows toward the cryocooler and the largest source of dissipated power occurs on the rod surface, the temperature reaches its maximum at the tip of the rod.
The iteration process continues until $e_{n}$ becomes sufficiently small [typically $e_n \sim{\cal O}(10^{-1})$]. Figure~\ref{fig:temp_final} shows typical iterations for temperature distributions along the rod. In the case when the cavity is coated with $Nb_{3}Sn$ the convergence is much faster compared with the case of pure niobium because the generated power is substantially lower for reasons explained below.

The convergence of the iteration process described above depends on dissipated power density, superconducting niobium thermal conductivity, and cryocooler operating temperature. The dissipated power density is directly related to the intensities of the electromagnetic fields that can be achieved inside the cavity and on stem geometry. The iterative procedure was repeated while altering the geometry of the cylindrical stem considered earlier. It was eventually found that a conically-shaped rod supports higher dissipation power while its surface remains at temperatures $<T_c$ thus enabling higher electromagnetic fields inside the cavity. 

\begin{figure}[h!]
   \centering
   \includegraphics[width=.95\columnwidth]{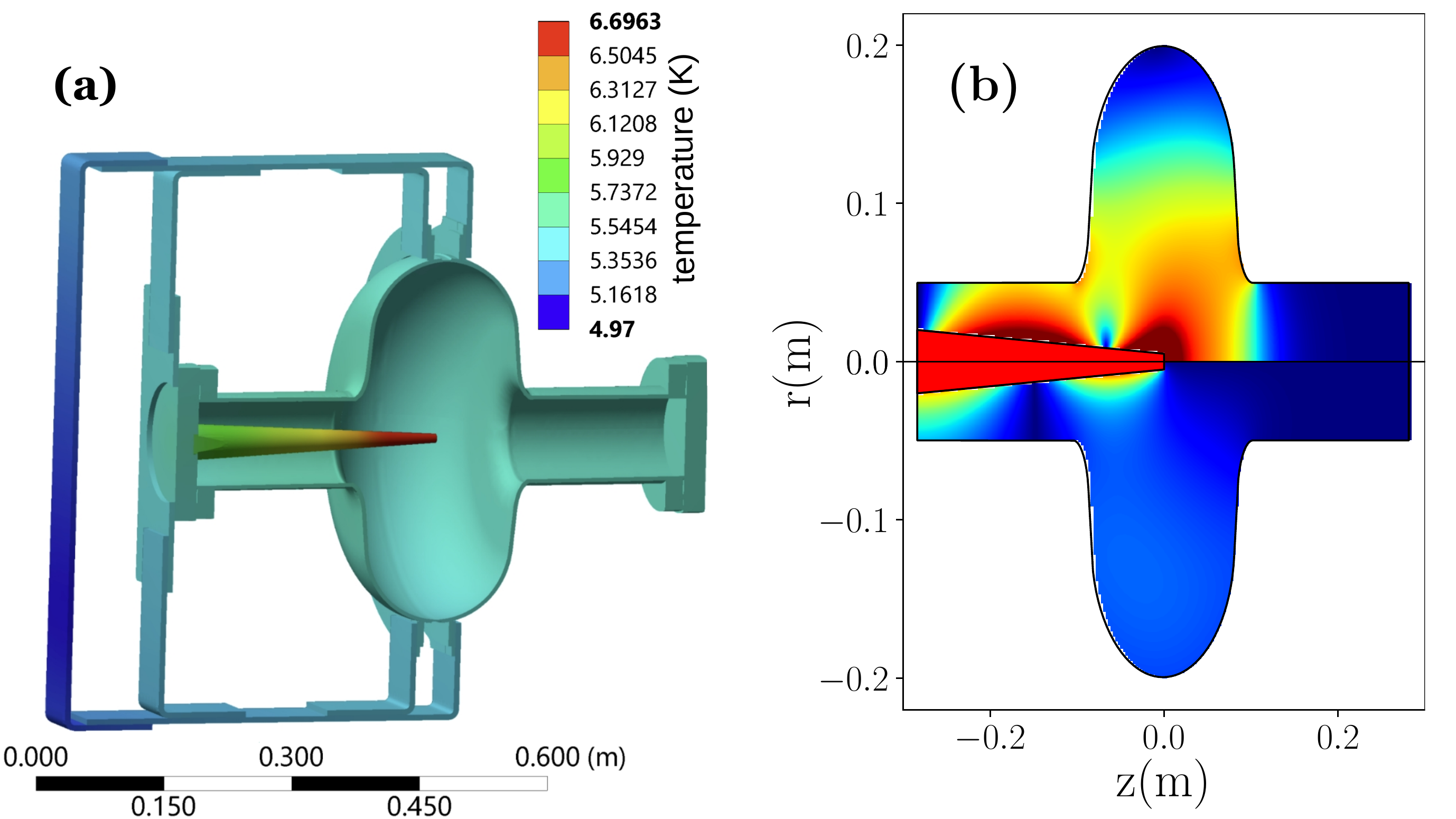}
   \caption{Steady-state temperature map (a) and associated peak-normalized electric- (upper plane $r>0$) and magnetic- (lower plane $r<0$)  field amplitudes (b) for the FE  configuration with optimized conical rod.}
   \label{fig:conical_final}
\end{figure}

Specifically, the devised geometry consists of a conical frustum rod with length 283~mm and radii of 5~mm and 20~mm at the tip and the base, respectively; see Fig.~\ref{fig:conical_final}(b). To avoid spurious field emission from sharp edges, the rod extremity has a rounded edge with a curvature of 0.5~mm. Such a configuration allows for a dissipated power of 1.0~W considering a cryocooler operating at a constant temperature of $T_k=3.4$~K. For the case of pure niobium the temperature at rod tip is 6.5~K; see Fig.~\ref{fig:conical_final}(a)  and the peak axial electric and surface magnetic fields are respectively 18.2~MV/m and 18.3~mT (below $B_{c1}= 170$~mT for Nb~\cite{Keckert_2019}). The corresponding cavity quality factor is estimated to be $Q\simeq 2.9 \times 10^{8}$. Additionally, the conical frustum rod does not appear to significantly alter the fields downstream of the cathode-surface location compared to the cylindrical-rod geometry as inferred by comparing Fig.~\ref{fig:conical_final}(b) and Fig.~\ref{fig:fieldpattern}(c). 

It is worth considering the case where the resonator shell is coated with a few \si{\micro\meter} thick layer of the Nb$_{3}$Sn superconductor. The alloy Nb$_{3}$Sn offers a lower surface resistance than pure Nb; see Fig.~\ref{fig:temp_final} (b). Typically, the ratio of surface resistance for these two materials at operating temperatures $\sim 5$~K is $R_{S,Nb}/R_{S,Nb_3Sn} \simeq 18$. Reproducing the previous thermal-electromagnetic simulations we found an Nb$_{3}$Sn coated-resonator with the same conical shape as previously devised to support a similar maximum temperature (6.8~K is simulated at the conical-rod extremity) as the one obtained for the case of Nb. The corresponding attainable peak electric and surface magnetic fields are 65.2~MV/m and 65.0~mT, respectively and the estimated quality factor is $Q\simeq 5.0 \times 10^{9}$. These values are significantly higher than the one obtained for pure niobium and in approximate agreement with the scaling described in Eq.~\ref{powerflux}. Such high field values have been demonstrated experimentally \cite{2020nb3sn}. Moreover,  Nb$_3$Sn offers the possibility to operate at even lower dissipated power while still maintaining the required low temperature. For instance, scaling the dissipated power by a factor $\simeq \left(\frac{31}{65}\right)^2$, i.e., from 1.0~W to 0.23~W would result in corresponding peak E and surface B fields of 31.1~MV/m and 31.0~mT ,respectively. operating at such low power would result in a final temperature of 3.9 K; see Fig.\ref{fig:temp_final} (a).

\section{Beam-dynamics considerations}
\begin{figure}[h!!!!!!!]
   \centering
   \includegraphics[width=1\columnwidth]{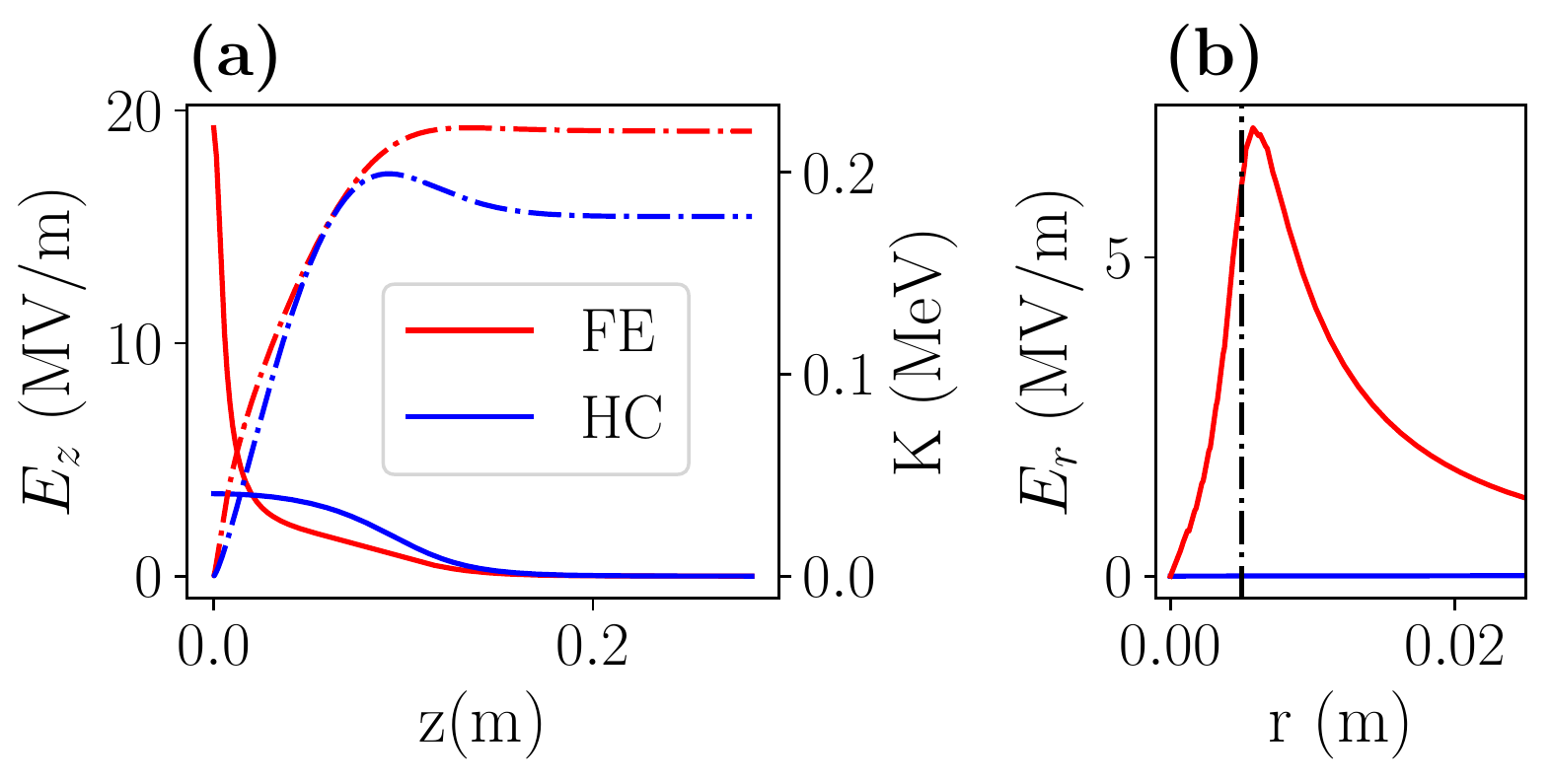}
   \caption{On-axis axial electric field $E_z(r=0)$ (solid traces) and kinetic energy $K$ (dashed lines) in  the FE (red) and HC (blue) configurations (a) and radial component of the electric field $E_r (z=\mbox{1.7~mm})$ (right). The dashed black line represent the radius of the rod (5 mm). The electromagnetic fields are normalized to the same dissipated power for comparison.}
   \label{fig:Energy}
\end{figure}

To fully assess the performance of the proposed configuration, we explore the generation of bright ultrafast electron beams via beam-dynamics simulations. We especially compare the performances associated with the proposed FE design with the beam parameters attainable in the commonly used HC configuration. The energy gain experienced by an electron propagating on axis of the two cavity configurations appears in Fig.~\ref{fig:Energy}(a) and confirms that despite the reduced accelerating gap, the FE geometry enables higher final energy (200~keV  compared to 140~keV for the HC geometry). Figure~\ref{fig:Energy}(b) also compares the transverse electric field on the cathode surface. A drawback from the FE geometry is the strong radial component of the electric field (close to the cavity-axis the  $E_r\simeq (-r/2)[\partial E_z(r=0,z)/\partial z]$) so that the conical-frustum rod results in a defocusing force at its tip. The radial field dependence on $r$ is more pronounced at radii approaching the edge ($r= 5$~mm) of the rod. 

The beam-dynamics simulations were performed using the particle-in-cell (PIC) program~\textsc{impact-t}~\cite{impact-t}. The program includes quasi-static space-charge effects using a mean-field algorithm and represents the electron bunch as an ensemble of macroparticles. We consider the case of photoemission where a laser pulse impinges the cathode to form an electron bunch. We assume the photo emitter to be deposited on the rod extremity. To quantify the brightness of the photo emitted electron bunch we introduce two metrics.  The first  one is the root-mean-square (RMS) transverse emittance defined as 
\begin{eqnarray}
\varepsilon_{\perp}=\frac{1}{mc}[\mean{x^2}\mean{p_x^2}-\mean{xp_x}^2]^{1/2},
\end{eqnarray}
where $m$ and $c$ are respectively the electronic mass and velocity of light, $(x,p_x)$ represents the horizontal position and momentum coordinates [the beam is taken to be cylindrical symmetric so that the emittance is the same for the horizontal $(x,p_x)$ and vertical $ (y,p_y)$ phase spaces), and $\mean{...}$ represents the statistical averaging over the macroparticle distribution. We also introduce the RMS bunch duration $\sigma_t$. The choice of these two figures of merit is rooted in the scaling of the bunch brightness as ${\cal B} \propto \frac{q}{\varepsilon_{\perp}^2 \sigma_t}$, where $q$ is the bunch charge. We specifically consider a beamline composed of the FE or HC resonator, a solenoidal magnetic lens followed by a drift space. The external fields (RF cavity and solenoidal lenses) are described by cylindrical-symmetric field maps. The solenoid lens controls the beam transverse emittance using a well-established compensation process~\cite{emit-comp}.  
\begin{figure}
    \centering
    \includegraphics[width=1\columnwidth]{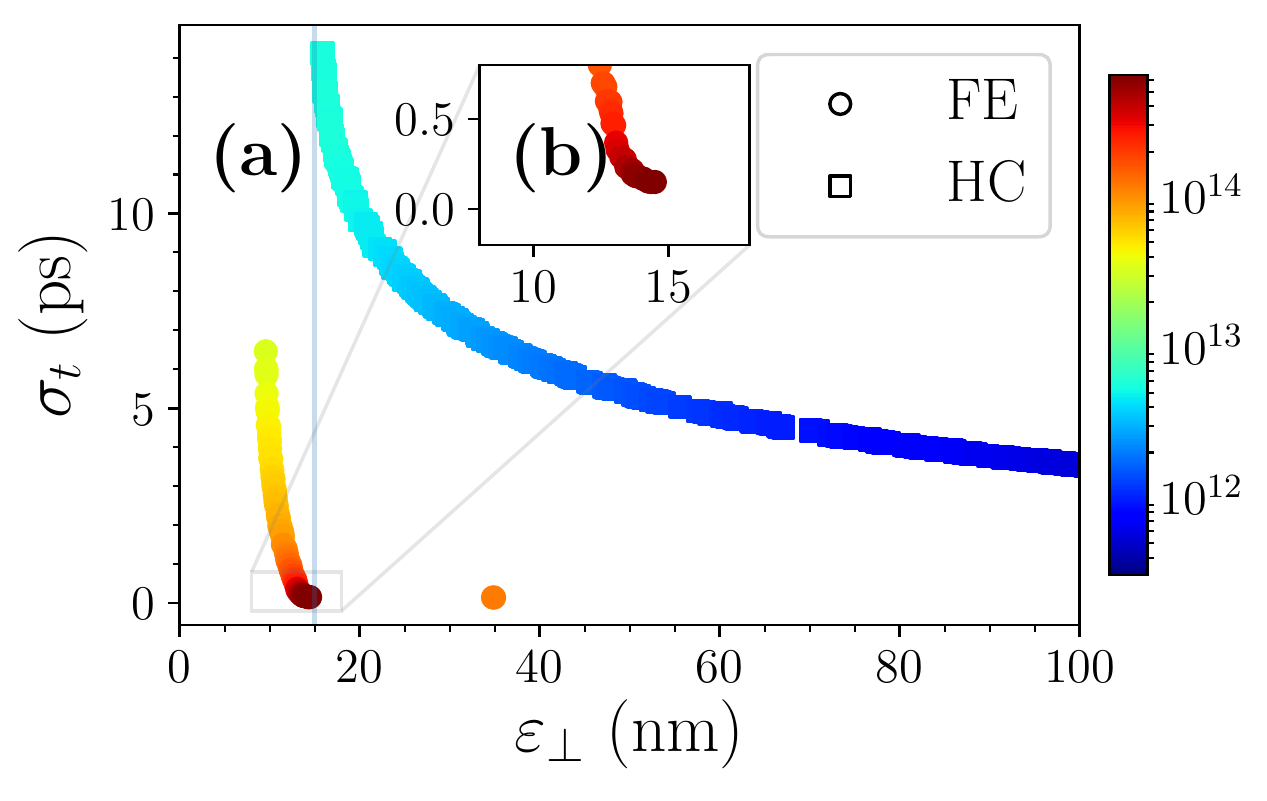}
    \caption{Pareto fronts between transverse emittance and bunch duration associated with the multi-objective optimizations for the HC and FE configurations. The colors represent beam brightness ${\cal B}$. The inset (b) displays a zoomed portion of the FE plot.}
    \label{fig:optResults}
\end{figure}
We assume the electron bunch to retain a cylindrical-symmetry throughout its transport through the system given the symmetry of the external fields. The configuration was optimized using the {\sc deap} multi-objective optimization framework~\cite{deap} to seek configuration that minimize both  $\varepsilon_{\perp}$ and $\sigma_t$. The control variables include the photoemission-laser spot size and pulse duration, laser injection phase w.r.t. the resonator field, the solenoid-lens location. The bunch charge was set to $q=20$~fC (corresponding to $1.25\times 10^5$ electrons). We use the optimized laser parameters to generate an initial macroparticle distribution at the cathode surface with an emittance corresponding to a mean-transverse energy of $180$~meV.  
The results of the optimization are summarized in Fig.~\ref{fig:optResults} where the Pareto fronts associated with the HC and FE geometries in the $(\sigma_t,\varepsilon_{\perp})$ space and confirms the proposed FE geometry consistently produce a beam with a factor $>10$ superior brightness compared to the HC geometry. The simulated beam parameters for the FE geometry are on par with those attained in state-of-the-art experiments at relativistic energies~\cite{maxson2017direct} while our setup is capable of operating in CW mode. 

\begin{figure}
    \centering
    \includegraphics[width=1\columnwidth]{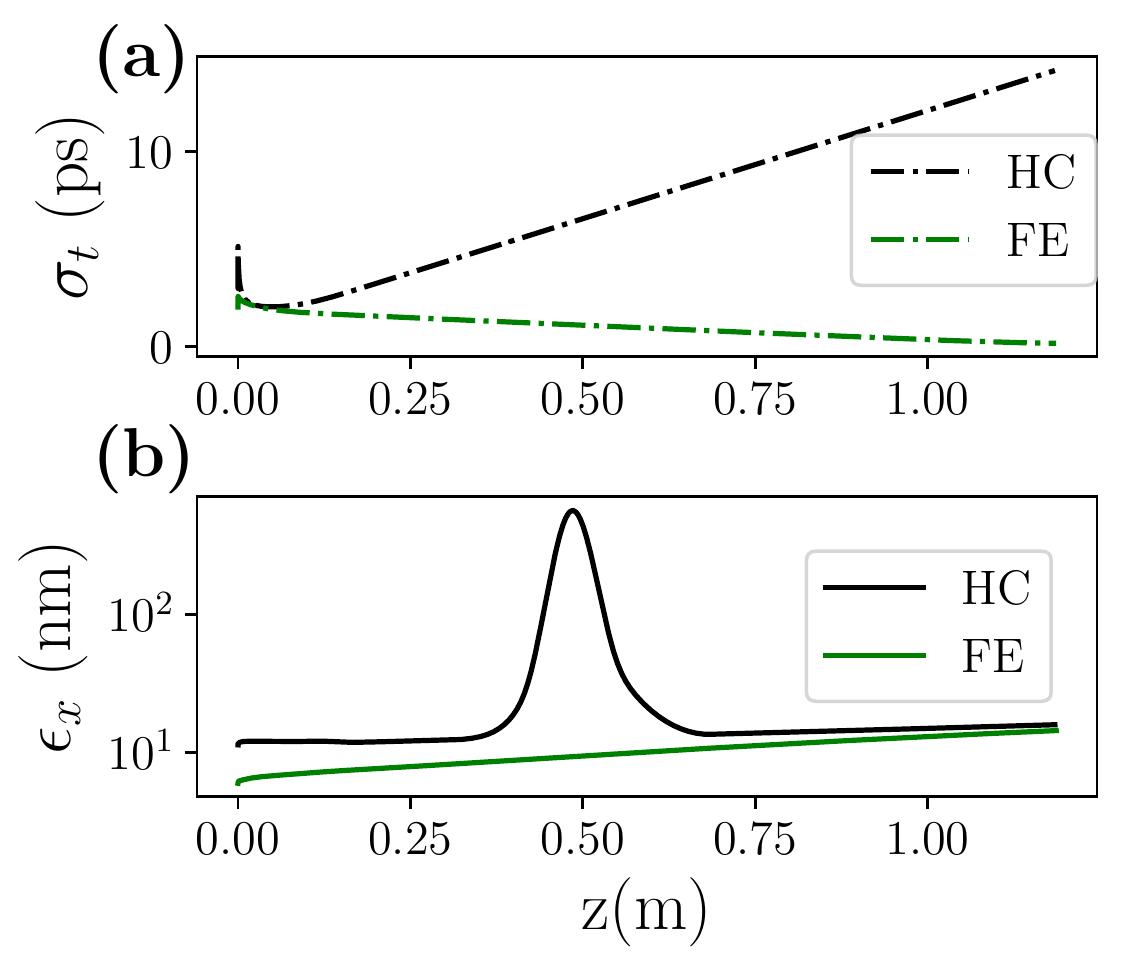}
    \caption{Evolution of the bunch duration $\sigma_t$ (a) and transverse emittance (b)  downstream of the HC (black) and FE (green) electron-source configurations. }
    \label{fig:beamLine}
\end{figure}

Figure~\ref{fig:beamLine} summarizes the evolution of the bunch duration and transverse  emittances along the optimized beamline for accelerator settings giving $\varepsilon_{
\perp}\simeq 15$~nm for both the HE and FC configurations; see Tab.~\ref{tab:summary}. The final bunch duration associated with the FE setup is one order of magnitude smaller than for the HC configuration.  
\begin{table}[t!!]
\centering
\caption{Beam line settings and simulated beam parameters downstream of the envisioned beam line .\label{tab:summary}}
\begin{ruledtabular}
\begin{tabular}{l c c c c }
\textbf{Parameter} & FE & HC & \textbf{unit}   \\
\hline
Laser pulse duration (rms) & 3.7 & 4.7  & \si{\pico\second}\\
Laser spot size (rms) &  10.00 & 20.00 &\si{\micro\meter}\\
Launch phase & 127.00 & 93.6 & deg \\

Final beam energy & 270 & 178.2 &\si{\kilo \eV} \\
Final transverse emittance (rms) & 15.0  & 16.0 &\si{\nano\meter\radian}\\
Final bunch duration (rms) & 0.17 & 14.14 & \si{\pico\second} \\
\end{tabular}
\end{ruledtabular}
\end{table}
\begin{figure}
    \centering
    \includegraphics[width=1\columnwidth]{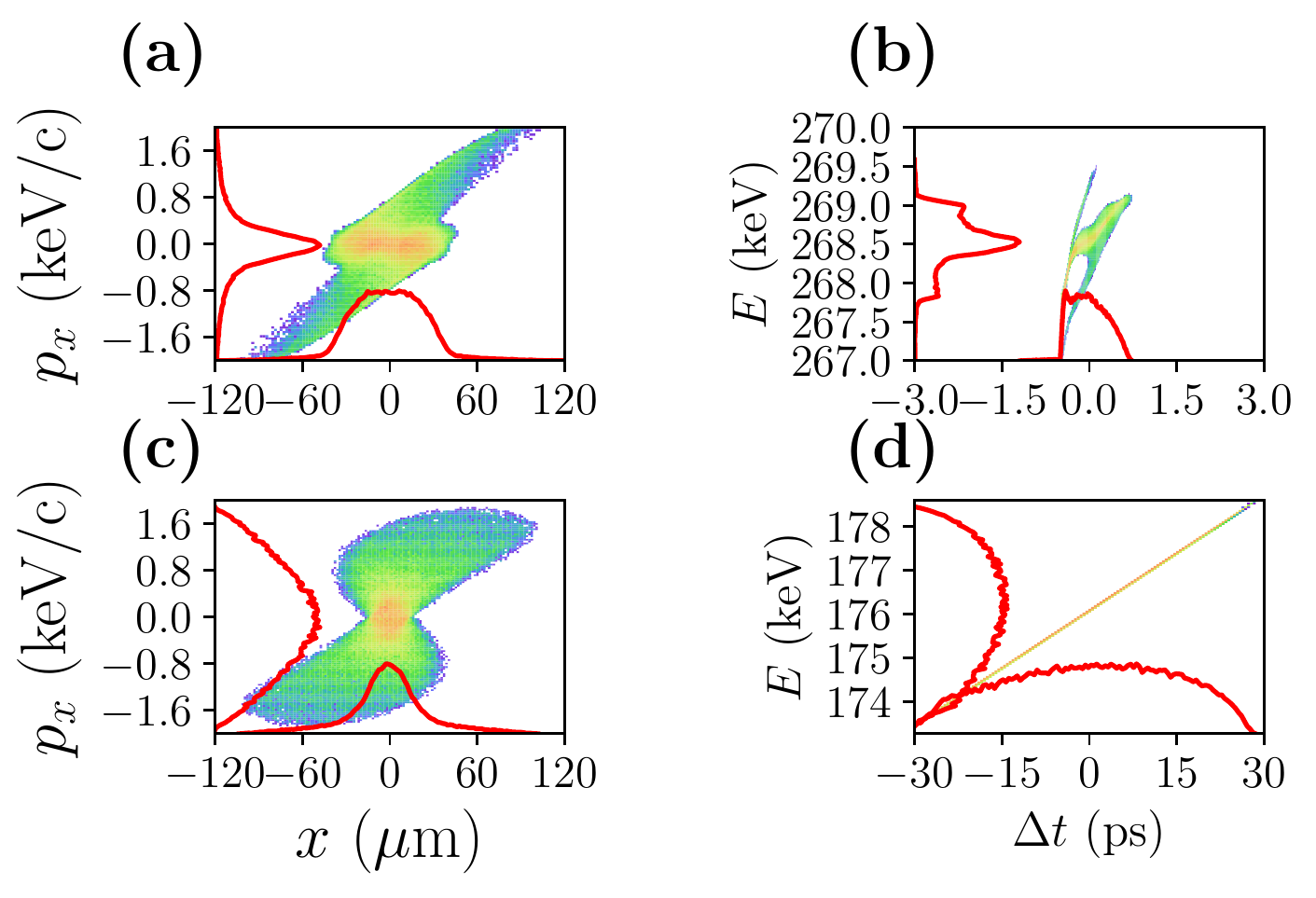}
    \caption{Transverse (a,c) and longitudinal (b,d) phase spaces distribution (density plots) for the FE (a,b) and HC (c,d) configurations. The traces represent the phase-space distribution projections along each of the direction.}
    \label{fig:phaseSpace}
\end{figure}
Finally, snapshots of the final transverse and longitudinal phase spaces appear in Fig.~\ref{fig:phaseSpace} for the two configurations. Due to the larger fields experienced by the bunch during the emission process in the FE configuration, the laser launch phase can be selected to form shorter bunches via the ballistic bunching without sacrificing the transverse emittance; see also Fig.~\ref{fig:beamLine}(a). The resulting longitudinal phase space associated with the FE configuration is at a longitudinal waist with an RMS bunch duration of  $\sim 170$~fs limited by third-order nonlinearities in the longitudinal phase space. This type of aberration could in principle be corrected following the technique discussed in Ref.~\cite{zeitler-2015-a}. 

\section{Conclusion}
In conclusion,  we demonstrated via numerical simulations that a single-cell elliptical resonator modified to enhance the field experienced by a bunch during emission can generate ultrafast electron bunches with superior transverse brightness than typically achieved in a standard half-cell configuration based on a similar elliptical geometry. The source design was optimized to be compatible with cryogen-free conduction cooling thereby resulting in a simple portable source. Nb$_3$Sn coating offers a substantial increase in the peak electric field while giving the possibility to operate at lower temperatures thus minimizing ohmic losses due to its lower $R_s$ when compared to pure Nb.

It should be stressed that important technical challenges remain before the practical realization of the concept especially regarding the possible cavity-performance degradation from the photoemitting material. Likewise, a removable cathode assembly would ultimately be preferable and would require the design of a Choke filter.  Nevertheless, the proposed setup could have interesting applications in ultrafast electron scattering experiments or could serve as an injector for optical-scale accelerator~\cite{DLA}. Although the energy is currently limited to 200~keV, the addition of another full-cell resonator downstream of the FE resonator thermally supported by a dedicated cryocooler could produce beams with MeV's energy. Ultimately, the proposed concept could be coupled with a field-emission source and open the path to MW average-power beams for a plethora of societal applications.

\section{Acknowledgements}
This work was supported by the US Department of Energy (DOE) under contract DE-SC0018367 with Northern Illinois University. Fermilab is managed by the Fermi Research Alliance, LLC for the DOE under contract number DE-AC02-07CH11359. We thank C. Ng (SLAC) and his team for providing access and guidance to the {\sc ace3p} suite of programs. This research used resources of the National Energy  Research  Scientific Computing Center which is supported by the Office of Science of the U.S. DOE, Contract No. DE-AC02-05CH11231.
\bibliography{ref}

\end{document}